\documentclass[entropy,review,accept,moreauthors,pdftex,10pt,a4paper]{mdpi} 
\usepackage{color}
\usepackage{soul}

\firstpage{1} 
\articlenumber{x}
\doinum{10.3390/------}
\pubvolume{18}
\pubyear{2016}
\copyrightyear{2016}
\externaleditor{Academic Editor: Ronnie Kosloff}
\history{Received: 4 October 2016; Accepted: 15 November 2016; Published: date}

\def \beq {\begin{equation}}
\def \edq {\end{equation}}
\def \ba {\begin{eqnarray}}
\def \ea {\end{eqnarray}}
\def \bes {\begin{subequations}}
\def \eds {\end{subequations}}
\def \beqn {\begin{equation*}}
\def \edqn {\end{equation*}}

\def \dag {\dagger}

\def \veps {\varepsilon}

\def \calh {{\cal{H}}}

\def \calg {{\cal{G}}}

\definecolor{misha}{rgb}{1,0,0}

\Title{Periodic Energy Transport and Entropy Production in Quantum Electronics}

\Author{Mar\'{\i}a Florencia Ludovico $^{1}$, Liliana Arrachea $^{1}$, Michael Moskalets $^{2}$, and David S\'anchez $^{3,}$*}
\AuthorNames{Mar\'{\i}a Florencia Ludovico, Liliana Arrachea, Michael Moskalets, and David S\'anchez}

\address{%
$^{1}$ \quad International Center for Advanced Studies, UNSAM, Campus Miguelete, 25 de Mayo y Francia, 1650~Buenos Aires, Argentina\\
$^{2}$ \quad Department of Metal and Semiconductor Physics,
NTU "Kharkiv Polytechnic Institute", \linebreak 61002 Kharkiv, Ukraine\\
$^{3}$ \quad Instituto de F\'{\i}sica Interdisciplinar y Sistemas Complejos
IFISC (UIB-CSIC), \linebreak E-07122 Palma de Mallorca, Spain
}

\corres{Correspondence: david.sanchez@uib.es; Tel.: +34-971-25-9887}

\abstract{The problem of time-dependent particle transport in quantum conductors is nowadays a well established topic.
In contrast, the way in which energy and heat flow in mesoscopic systems subjected to dynamical drivings
is a relatively new subject that cross-fertilize both fundamental developments of quantum thermodynamics
and practical applications in nanoelectronics and quantum information. In this short review, we discuss
from a thermodynamical perspective recent investigations on nonstationary heat and work generated in quantum systems,
emphasizing open questions and unsolved issues.}

\keyword{quantum thermodynamics; time-dependent heat; AC-driven quantum systems}

\begin{document}



\section{Introduction}

The rapid progress in the manipulation of small systems at submicron scales has spurred
the interest in the field of quantum thermodynamics. The goals of this discipline is the understanding of the fundamentals
of thermodynamics in quantum systems typically driven out of equilibrium by means of external fields.
Most of the recent literature focuses on static fields  and the resulting stationary transport effects
(for a recent review, see~\cite{ben16}).
However, there is a growing interest in analyzing the thermodynamic properties of quantum conductors
in the presence of time-dependent potentials. In this case, dynamics is the main objective of the theory 
as fluxes and responses depend explicitly on time~\cite{kos13}.

Quantum electronic devices are paradigmatic realizations of open quantum systems. Typically, we can identify in these setups a finite-size piece where a small number of electrons are confined. However, these particles are not isolated. They are in contact with leads, substrates and the electromagnetic environment, which in practice constitute  macroscopic thermal and particle reservoirs. We can distinguish two main classes of time-dependent processes in these systems. One of them is related to the transient behavior after a short-time non-equilibrium perturbation, where the main issue is to understand the relaxation and thermalization processes as the system approaches the equilibrium state. The other class is the response to a periodic driving. Both classes of dynamics have received a significant interest for some years now. 

From the conceptual point of view, it is quite natural to relate the energy transport in periodically-driven systems to thermodynamic machines, like heat engines and refrigerators, since these typically operate in cycles. There is a lengthy literature on nanoscale quantum thermodynamics based on two-level systems \cite{twolevel1,twolevel2,twolevel3,twolevel4,twolevel5}, atomic \cite{atomic1,atomic2,atomic3,atomic4} and molecular
\cite{molecular1,molecular2,molecular3} systems, nanomechanical \cite{nanomechanical1,nanomechanical2,nanomechanical3,nanomechanical4,nanomechanical5,nanomechanical6} systems and other systems represented by harmonic oscillators \cite{osci1,osci2,osci3,osci4}. The list of works devoted to study cyclic processes in electron systems is shorter but is growing every week.  In~\cite{eng} heat transport induced by adiabatic pumping is analyzed in a quantum dot driven by two AC voltages operating with a phase lag. Two interesting operational modes were identified: (i) the heat pumping against a gradient of temperature and (ii) the exchange of work between the driving forces. Heat pumping by AC driving was also investigated in the non-adiabatic regime \cite{rey} and in arrays of quantum dots in the Coulomb blockade regime \cite{janine}, while in \cite{mos-bu2009} the effect of the exchange of power between two driven quantum capacitors was analyzed. In  \cite{bus}
the exchange of power between electron currents induced by DC bias voltages and adiabatic driving forces was considered and a similar mechanism was later analyzed in the context of a pair of helical edge states coupled to a magnetic island \cite{torque}. The effect of AC driving in the efficiency of usual two-terminal thermoelectric setups was addressed in  \cite{janine,dar16,zho15}. An extended  thermoelectric framework to describe the work exchanged with the AC driving sources within the adiabatic regime, on equal footing with the exchange of charge and heat between electron reservoirs with thermal and voltage biases was proposed in  \cite{lud16}. Thermal transport in electron systems driven by the coupling to electromagnetic radiation has been studied in \cite{Bergenfeldt:2014he,heat-phot,heat-phot1,heat-phot2}.
The related problems of heat fluctuations and the definition of effective temperatures to characterize the net heat transport in these systems were addressed in~\cite{Kindermann:2004im,Averin:2010kp,Sergi:2011eo,mos14,bat14,san13,bat14b,cre15,cre16,eym16,heat-eng3} and \cite{temp1,temp2,temp3}, respectively.

In addition to the interest in the fundamental aspects, it is of paramount importance for potential applications to discriminate which portion of the energy invested to operate these devices is amenable to be used and which one is wasted by dissipation. This distinction is at the heart of thermodynamics and is conventionally addressed in quasi-stationary processes where the system under study is very weakly coupled to the reservoirs. In quantum electronics, however, the generic situation is to have the driven structure strongly coupled to the rest of the circuit, which plays the role of the reservoir.
On the other hand, we are typically interested in the generation of currents, which implies  non-equilibrium situations.

In this paper, we review recent developments in dynamical quantum thermodynamics applied to nanoelectronic systems.
We will discuss in detail the energy transfer in out-of-equilibrium systems coupled to external baths.
The analysis is based on a phase coherent mesoscopic sample (a confined system with discrete energy levels)
attached to fermionic reservoirs held at a given temperature. The energies of the sample evolve with time
due to the coupling with nearby AC gate terminals. 
Deep inside the reservoirs, electrons relax their excess energy, and the baths can thus be considered
in local thermal equilibrium. We will also consider the entropy production in the whole system and
will identify the different terms arising in the redistributed energy and heat~\cite{lud14,lud16b}.
Importantly, when the energies shift slowly with time, the response is adiabatic, and an exact Joule law can be demonstrated for the time domain~\cite{lud14,lud16b,mh16}. 
In addition, we will briefly discuss the nonadiabatic regime~\cite{lud14b}
in which case the AC frequency is larger or of the order of the inverse dwell time inside the conductor. Our analysis is completely general and does not rely on the particular approach followed to
evaluate the relevant dynamical quantities, like the charge and energy fluxes and the rate of entropy production. 

Dynamical transport in quantum conductors is a relatively old subject with
established theoretical principles~\cite{but93,brou,pump1,pump2}, 
which has been investigated with different techniques~\cite{past,jauho,pla04,koh05}. It is well understood that a Floquet scattering matrix approach (FSM)~\cite{mos12} can account
for photon-assisted phenomena in nanostructures (absolute negative conductance and dynamical localization~\cite{pla04}, single-electron emission \cite{mos08,mh16}, etc.). In the present work, we also include for completeness the details of the calculation of these quantities with the Keldysh formalism using the
Floquet-Fourier representation as in~\cite{arrg1,arrg2} and discuss the connection to the FSM formulation as in~\cite{arr-mos}.

\section{Theoretical Model}

\begin{figure}[ht!]
\centering
\includegraphics[width=7cm]{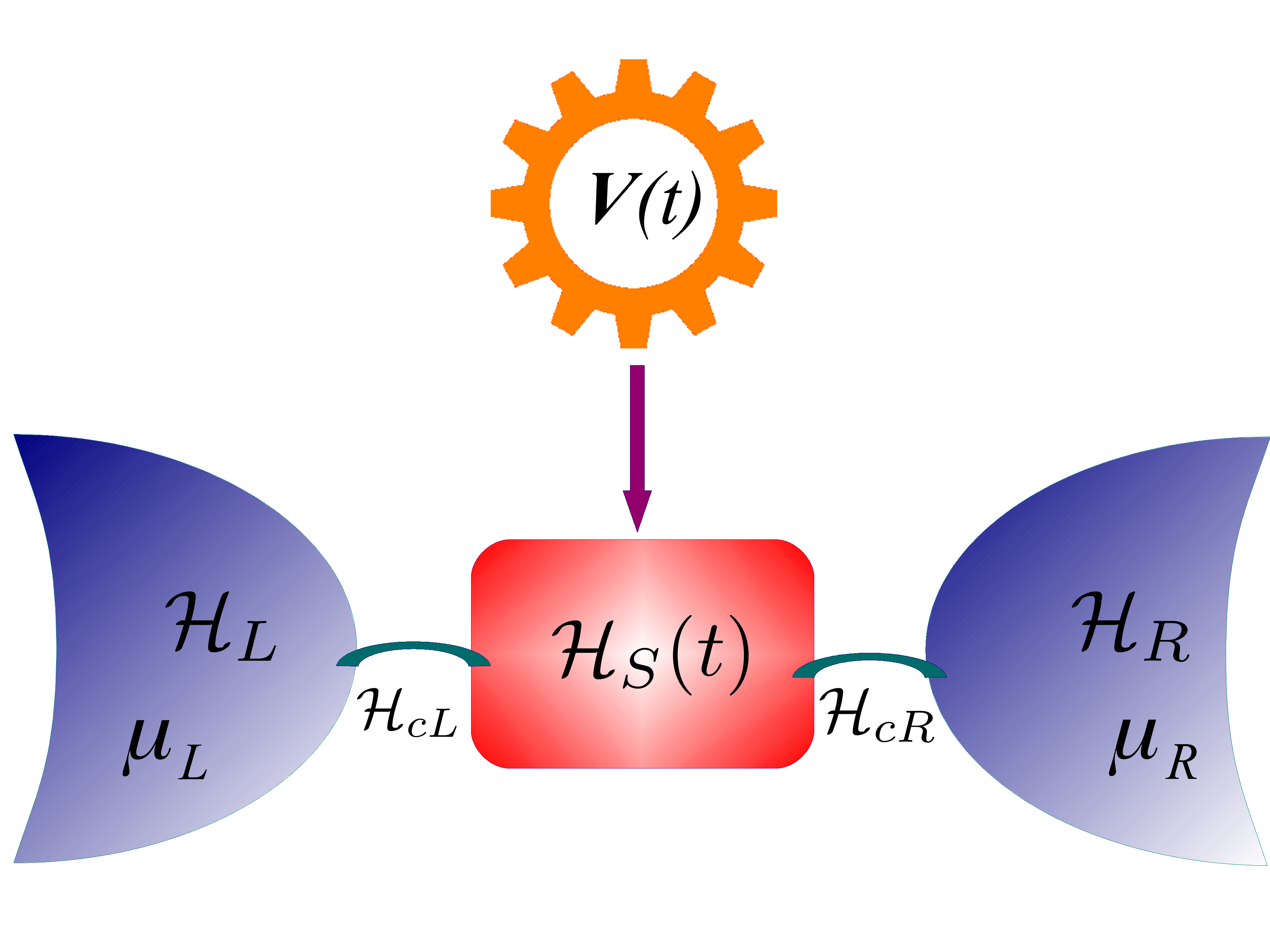}
\caption{Sketch of the system under consideration. A quantum conductor (described by the Hamiltonian $H_S$), is coupled to two reservoirs ($H_L$ and $H_R$) kept at the same temperature $T$, but with different chemical potentials $\mu_L$ and $\mu_R$. The conductor is also driven out of equilibrium by the application of AC local power sources, which are all collected in the vector ${\bf{V}}(t)$. The Hamiltonians representing the left and right contact regions are $H_{cL}$ and $H_{cR}$, respectively.
\label{fig1}}
\end{figure}

We consider that our system is a slowly-driven (adiabatically)  quantum conductor (e.g., a quantum dot or an array of quantum dots) coupled to several baths (fermionic reservoirs), see Figure~\ref{fig1} for the two terminal case.
In the adiabatic regime, it is possible to propose a thermodynamic function within an extended resonant level model that satisfies both the first and the second laws~\cite{lud14,lud16b}.
We do not make any assumption on the coupling strength between the system and the baths. 

Note that the problem of defining a system and a bath in the strong coupling regime has been addressed recently~\cite{bru16}. 
Moreover, the authors of~\cite{och16} claim that the extended resonant level model is unable to properly describe energy fluctuations despite the fact that the expectation value of the energy is correctly derived. 
On the other hand,~\cite{car16} finds that an influence functional approach~\cite{car15,twolevel4} applied to the spin-boson model can yield exact expressions for the energy dissipation in a driven open quantum system strongly coupled to a heat bath. 

Here, we take into account the Hamiltonian of the fermionic baths,
$\calh_{\rm res}= \sum_{\alpha} \calh_{\alpha} $ with 
$ \calh_{\alpha}= \sum_{ k_{\alpha}}\veps_{k_{\alpha}}c_{k_{\alpha}}^{\dag}c_{k_{\alpha}}$,
where $\veps_{k_{\alpha}}$ is the energy dispersion relation and $c_{k_{\alpha}}^{\dag}$ ($c_{k_{\alpha}}$)
creates (destroys) an electron with wavenumber $k_{\alpha}$,
the Hamiltonian of the system, $\calh_S( {\bf V}(t) )$, where the functions ${\bf V}(t) ={\bf V}(t+\tau)= (V_1(t), \ldots, V_M(t))$ 
describe the AC driving power sources capacitively coupled to the system,
$\tau$ being the AC period, and the Hamiltonian of the contact region between the system and the baths,
$\calh_{\rm cont} =  \sum_{\alpha} \calh_{c \alpha}$ with $
\calh_{c \alpha}= \sum_{k_{\alpha}}\left( w_{k_{\alpha}} c^{\dagger}_{k_{\alpha}} d_{l_{\alpha}} + H. c \right)$,
where $w_{k_{\alpha}}$ are tunneling amplitudes, $l_\alpha$ denotes the site of the central part to which the reservoir $\alpha$ is attached,
and $d_{l_{\alpha}}$, and  $d^{\dagger}_{l_{\alpha}}$ destroys and creates, respectively,
an electron in the central system. The three contributions add up to the Hamiltonian
of the full system,
\beq\label{eq_H}
\calh(t) = \calh_{\rm res} +\calh_S(t)+ \calh_{\rm cont} \,,
\edq 
which explicitly depends on time due to the applied AC potentials.
We stress that the explicit form of $\calh_S$ remains unspecified, because we will below present a thermodynamic
discussion that is completely general and independent of the sample details. In fact, our presentation
is also valid for interacting electrons provided these carriers are confined to the mesoscopic region.
The fermionic baths are metallic leads with good screening properties and electron-electron interactions
can be safely neglected there. 
The latter is an important remark since a gauge-invariant transport theory in the AC regime must include the self-consistent Coulomb interaction~\cite{but94}. 
This requirement in the case of the heat current was discussed in~\cite{che15}.

Detailed knowledge of the heat dynamics in quantum electronics is also relevant
for time-dependent thermoelectrics, where temperatures may be modulated with time.
A calculation for quantum $RC$ circuits shows~\cite{lim13}
that thermoelectric charge relaxation resistance significantly differs from its
purely electric counterpart. 
In a different setup, the thermoelectric power factor has been experimentally demonstrated
to grow in the photo-induced regime~\cite{lv15}.
When the coupling between a dot and its attached leads quickly changes with time,
transient currents are predicted to occur in the absence of tunneling~\cite{dar16}.
These findings call for a unified description of thermoelectrics and heat for AC-driven
conductors. This is accomplished in~\cite{lud16}, where the authors generalize the
Onsager reciprocity relations for the case of adiabatic drivings.
Thermodynamic forces and fluxes can be identified this way~\cite{pro16}.
Unfortunately, experimental measurements of thermoelectric responses in the time domain are scarce.
An exception is~\cite{gar16}, where a frequency-resolved impedance spectroscopy
method is used in thermoelectric modules.

\section{First Law}

Let $\rho= e^{-\beta( {\cal H}-\mu {\cal N})}/Z$ be the density matrix,
with $Z= \mbox{Tr}\left[  e^{-\beta( {\cal H}-\mu {\cal N})} \right]$ the partition function, ${\cal N}$ the particle number and $\beta=1/(k_BT)$.
We start by considering $\mu_L=\mu=R=\mu$ and $T_L=T_R=T$.
 Variations $\delta \rho= \rho(t+\delta t) - \rho(t)$
are due to changes in $\delta {\bf V}={\bf V}(t+\delta t)-{\bf V}(t)$,
which we assume to be small. Then, both the particle number and internal energy
change according to:
\beq \label{dn}
\delta N_{\nu} =  \mbox{Tr}\left[  \delta \rho {\cal  N}_{\nu} \right],\; \nu=\alpha, S
\edq
and:
\beq \label{du}
 \delta U_{\nu} =   \mbox{Tr}\left[  \delta \rho {\cal H}_{\nu} \right],\; \nu=\alpha, c\alpha, S
 \edq
respectively. The particle number and energy for the full system remain constant, since the system plus the baths can be considered as a larger, closed system.
\beq
\delta N = \sum_{\alpha} \delta N_{\alpha} + \delta N_S  \,,
\edq
and
\beq
 \delta U =  \mbox{Tr}\left[  \delta \rho {\cal H} \right]= \sum_{\alpha}\left[ \delta U_{\alpha} + \delta U_{c \alpha} \right] + \delta {U}_{S}  ,
 \edq

The first law of thermodynamics states that total heat in the system is:
\beq \label{delqt}
\delta Q_{\rm tot}\vert_{\mu,T} = \sum_{\alpha}\left[ \delta U_{\alpha} + \delta U_{c \alpha} - \mu \delta N_{\alpha}  \right] + \delta {U}_{S} - \delta W_{\rm AC}  - \mu \delta N_S\,,
\edq
where:
\beq
\delta W_{\rm AC}=- \mbox{Tr}[  \rho \partial {\cal H}/\partial {\bf V} ] \delta {\bf V}
\edq
is the work developed by the time-dependent driving sources. The corresponding change in the entropy reads:
\beq \label{delst} 
T \delta S \vert_{\mu,T} =  \delta Q_{\rm tot}\vert_{\mu,T} =  -  \delta W_{\rm AC},  
\edq
where we have used in the last equality the conservation  of particles and energy expressed by  Equations (\ref{dn}) and (\ref{du}).

We now consider a more general situation, where the reservoirs have slightly different chemical potentials $\mu_{\alpha}= \mu + \delta \mu_{\alpha}$.
 This implies investing an extra electric work done by DC bias voltages 
\beq
\delta W_{\rm el}= \sum_{\alpha} \delta \mu_{\alpha} \delta N_{\alpha}.
\edq
Here, $\delta \mu_{\alpha}$ is a small departure of the 
electrochemical potential of lead $\alpha$ with respect to the common chemical potential $\mu$.

Equation~(\ref{delst}) generalizes to:
\beq \label{delstg} 
 T \delta S \vert_{\mu,T} + T  \delta S \vert_{\delta \mu} =   \delta Q_{\rm tot}
\edq
with
\beq
T \delta S \vert_{\delta \mu} = -  \sum_{\alpha} \delta \mu_{\alpha} \delta N_{\alpha} = - \delta W_{\rm el}.
\edq

Since $\delta N=\delta U=0$ due to particle and energy conservation, the first law
implies that the total heat generated in the full system stems from 
the work done by the AC and the DC sources,
\beq
\delta Q_{\rm tot} = - \delta W_{\rm AC}- \delta W_{\rm el} .
\edq

\section{Heat Current and Power}
The rate of change for the energy corresponding to the different parts of the system
(sample $S$, reservoir $\alpha$ or coupling region $c\alpha$)
is given by the exact quantum-mechanical expression
\beq \label{encur}
J^E_{\nu}(t) =  \frac{i}{\hbar} \langle \left[ {\calh}, {\calh}_{\nu} \right] \rangle \,,
\edq
where $\nu \equiv S, \alpha, c \alpha$. The time variation of the expected value
of the total Hamiltonian ${\calh}$,
\beq \label{enbal}
\dot{  \langle \calh \rangle }= \sum_{\alpha} \left[  J^E_{\alpha}(t) + J^E_{c \alpha}(t) \right] + J^E_S(t) - 
{\bf F} \cdot \dot{\bf V}\,,
\edq
is given by the explicit time derivative of $\langle{\calh}\rangle$
in terms of the generalized force ${\bf F} = - \langle \partial \calh/\partial {\bf V} \rangle$.
On the other hand, we know that the change in time of the total energy stored in the full setup, containing the central system, contacts and reservoirs, is equal to the total power developed by the external AC sources, $\dot{  \langle \calh \rangle }=-P_{\rm AC}(t) $. The latter reads:
\beq
P_{\rm AC}(t)={\bf F}\cdot \dot{\bf V}.
\edq
Hence, it follows that the
fluxes obey:
\beq \label{intcon}
\sum_{\alpha} \left[  J^E_{\alpha}(t) + J^E_{c \alpha}(t) \right] + J^E_S(t) =0\,.
\edq

The time variation of the charges present in the different pieces of the system
can be similarly inferred. We start from:
\beq\label{charcur}
I_{\nu}^C(t)=e\dot{\langle {\cal N}_{\nu} \rangle }=
 \frac{ie}{\hbar}\langle \left[ {\calh}, {\cal N}_{\nu} \right] \rangle,\edq
with $\nu=\alpha,S$.
Conservation of the total charge demands that the currents fulfill:
 \beq \label{numbal}
e\dot{ \langle {\cal N} \rangle } =  I^C_S(t)+\sum_{\alpha} I^C_{\alpha}(t)=0.
\edq 
Noticeably, the coupling region contributes to the energy current balance [Equation~\eqref{intcon}]
but not to its particle counterpart [Equation~\eqref{numbal}]. This will have important
consequences for the heat current definition, as explained below.

Using Equation~\eqref{delqt} it is straightforward to derive the total heat flux,
\ba
 \dot{Q}_{\rm tot}(t) &=& \sum_{\alpha}  \left[J^E_{\alpha}(t)- \mu_{\alpha} \frac{I^C_{\alpha}(t)}{e} +J^E_{c\alpha}(t) \right]+ J^E_S(t)  \nonumber \\
 & &- P_{\rm AC}(t) -\mu \frac{I^C_S(t)}{e}.
 \ea
 Now, we define:
 \beq\label{powerbat}
 P_{\rm el}(t) =\sum_{\alpha} \delta \mu_{\alpha} I^C_{\alpha}(t)/e,
 \edq
 which is the electric power developed by the batteries applied to the fermionic baths to keep the electron flow. 
 It follows from Equations~(\ref{intcon}) and~(\ref{numbal}) that 
 \beq
 \dot{Q}_{\rm tot}(t) = -P_{\rm AC}(t) - P_{\rm el}(t),
 \edq
 which states that the total heat flux in the system is produced by the AC and DC driving powers, as expected.
 
The question now is how to define physically-meaningful time-dependent currents flowing through different pieces of the setup. 
 The work in~\cite{lud14} demonstrates that
the conventional definition for the heat flowing in the reservoir $\alpha$,
\beq\label{defheatres2}
 \dot{\tilde{Q}}_{\alpha}(t)= J^E_{\alpha}(t) - \mu_{\alpha} \frac{I^C_{\alpha}(t)}{e},
\edq
suffers from serious drawbacks. In particular, for a slowly-driven quantum level coupled
to a single reservoir the use of $\dot{\tilde{Q}}$ leads to negative heat fluxes at zero
temperature. This would imply that heat would be extracted from a reservoir
at zero temperature, which makes no sense if we are to interpret this heat as dissipation.
In contrast, when one considers the expression
\beq\label{defheatres}
 \dot{Q}_{\alpha}(t)= J^E_{\alpha}(t) + \frac{J^E_{c\alpha}(t)}{2} - \mu_{\alpha} \frac{I^C_{\alpha}(t)}{e}.
\edq
one finds that the heat flux is always positive at zero temperature, i.e.,
energy enters into the reservoir and is evacuated from the power source, as should be. Moreover, Equation~(\ref{defheatres}) is a sound expression
that fulfills four nice properties:
\begin{itemize}[leftmargin=*,labelsep=4mm]
\item	It leads at low frequencies to a correct Joule law valid for all times~\cite{lud14}.
\item	It shows perfect agreement between the Green function approach and the scattering matrix formalism~\cite{lud14}.
\item	It displays parity symmetry upon reversal of the AC frequency even for interacting quantum conductors~\cite{ros15}.
\item    It reduces to the conventional definition in the stationary case since the term $J^E_{c\alpha}(t)/2$ vanishes after time averaging~\cite{lud16b}.
\end{itemize}

However,~\cite{esp15,esp15b} propose not to use any heat flux definition due to
apparent contradictions with the third law of thermodynamics and the onset
of thermodynamic inconsistencies. We would like to clarify and highlight that those references consider a very different setup, in which the time dependence of the Hamiltonian also takes place in the contact region (in our case only the central part of the system evolves in time). This might seem as a minor difference, but on the contrary it is a fundamental point, because in that case there is a part of the energy stored in the contact region that should be interpreted as dissipative power and hence it will not contribute to the heat flux.

\begin{figure}[ht!]
\centering
\includegraphics[width=6cm]{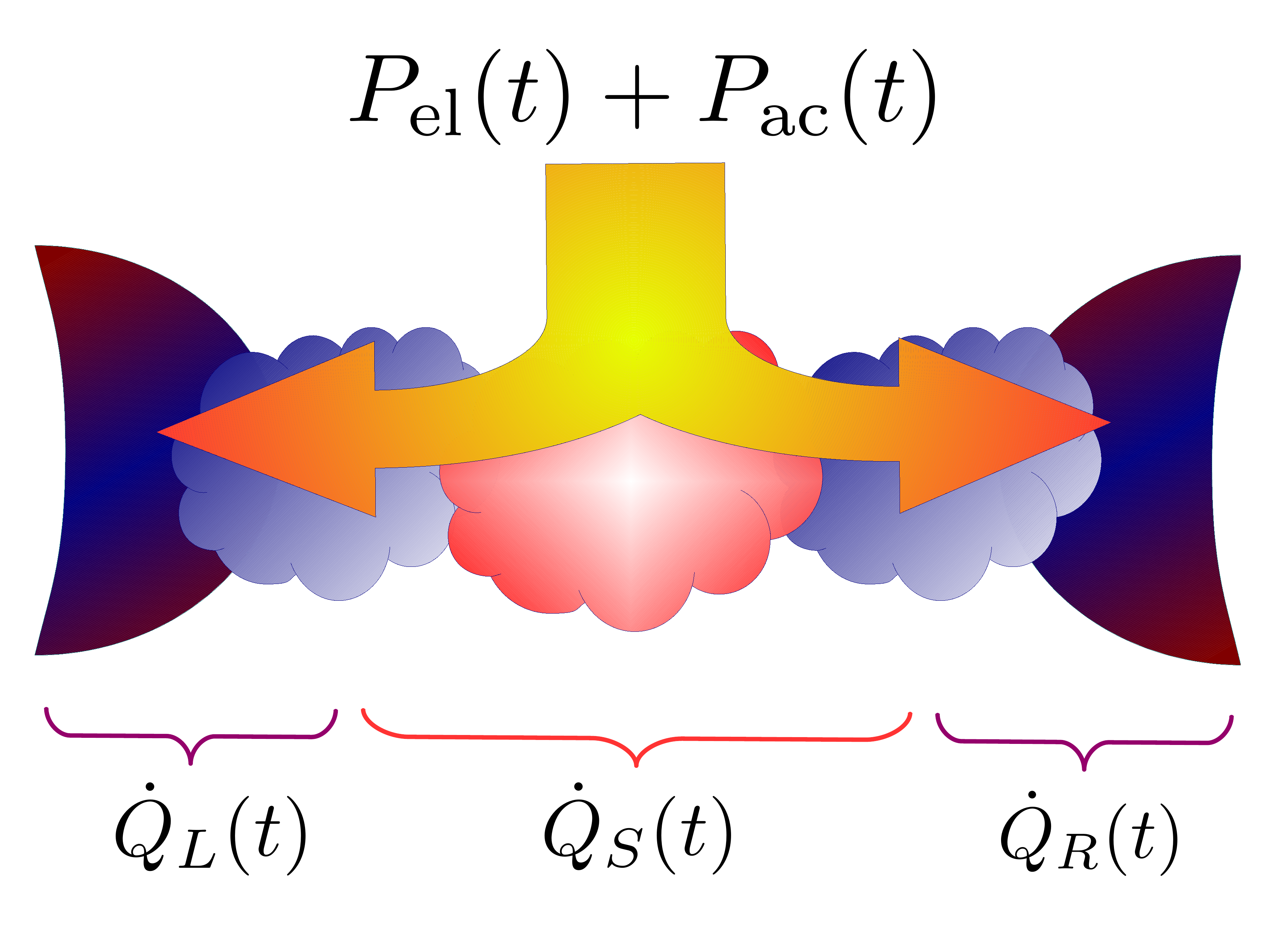}
\caption{The total heat flux in the system is produced by the AC and DC driving sources, which is summarized by the relation $\dot{Q}_{\rm tot}(t)=-P_{\rm AC}(t)-P_{\rm el}(t)$. The definition of the heat currents entering the reservoirs, as well as the heat variation in the central part, takes into account half of the energy stored in the contact regions. The heat flux of the central piece contributes purely dynamically, since it is zero when averaged over one period of the AC driving potentials. In the stationary state, charge and energy can be stored only at the reservoirs.
\label{fig2}}
\end{figure}

We emphasize that the extra term $J^E_{c\alpha}(t)/2$ has a pure dynamical origin, since it vanishes when it is averaged over a period. This makes sense from the physical point of view, because in a period the energy can be stored only at the reservoirs.  
On the other hand, this contribution of the contact region is also unique to the heat current. For example, 
the particle flux is independent of the charge variation in the coupling region, simply because it does no exist.
In stark contrast, the tunneling term does contribute to the energy transfer between the quantum system and the bath
and must be taken into account in any thermodynamically consistent calculation of time-dependent heat currents.
Analogously, we can define the heat variation in the sample as
\beq\label{defheats}
 \dot{Q}_{S}(t)= \dot{E}_S(t) - \mu \frac{I^C_{S}(t)}{e}+  \sum_{\alpha} \frac{J^E_{c\alpha}(t)}{2} .
\edq
such that the total heat satisfies $\dot{Q}_{\rm tot}(t)= \sum_{\alpha} \dot{Q}_{\alpha}(t) +  \dot{Q}_{S}(t)$, see Figure~\ref{fig2}.

\section{Entropy Production}

Further insight can be gained if we examine the instantaneous entropy production, which is presumably a meaningful quantity for low driving frequencies, for which the system appears to be almost stationary and the classical laws of thermodynamics remain valid. Otherwise, for high enough frequencies, the nonequilibrium and quantum effects may be significant, which may hinder the use of the classical thermodynamics. This regime requires further investigation.
The total entropy production has been demonstrated to follow an integral fluctuation
theorem for a driven system strongly coupled to a heat bath~\cite{sei16}.
Here, we stress that the entropy production associated with dissipative processes is always positive, in agreement with the second law. We begin by noticing that the AC forces have, in general, conservative (cons) and nonconservative (nc) components. In fact, in~\cite{lud16} it was shown that ${\bf F}(t)= {\bf F}_t + 
{\bf F}^{\rm nc}(t)$, where the first term corresponds to the Born-Oppenheimer component, which is evaluated with the equilibrium density matrix $\rho_t$ corresponding to the frozen Hamiltonian at the time $t$ and ${\bf F}^{\rm nc}(t)$ collects the nonequilibrium components.
The first term corresponds to an evolution given by a sequence of equilibrium states which are, thus, reversible and do not contribute to the entropy production. Hence \linebreak $P_{\rm AC}(t)=P^{\rm cons}(t)+
P^{\rm nc}_{\rm AC}(t)$. The electric power of the batteries is purely non-conservative (dissipative).
In this way, the 
total power can be split into dissipative and conservative forces
$P_{\rm tot}(t)=P_{\rm el}(t)+ P_{\rm AC}(t)= P^{\rm diss}_{\rm tot}(t)+  P^{\rm cons}(t)$, where $ P^{\rm diss}_{\rm tot}(t)=
P_{\rm el}(t)+P^{\rm nc}_{\rm AC}(t)$.

As a result, the total entropy production reads,
\beq \label{2law-1}
\dot{S}^{\rm diss}(t)= \frac{1}{T}\left[\sum_{\alpha}  \dot{Q}_{\alpha}(t) + \dot{Q}^{\rm diss}_S(t) \right] = -\frac{1}{T} P^{\rm diss}_{\rm tot}(t)\,.
\edq
We have defined $\dot{Q}^{\rm diss}_S(t) = \dot{Q}_S(t) -P^{\rm cons}(t)$, i.e., we identify the irreversible heat production at the central system as the result of subtracting the power developed by the conservative AC forces to the total heat instantaneously stored by the
electrons that are in direct contact with the AC sources. 
After averaging over an AC cycle, the conservative part of the heat current vanishes.
Furthermore, the heat variation in the sample also vanishes,
$\overline{\dot{Q}^{\rm diss}_S}=\overline{\dot{Q}_S}=0$.
The remaining terms have nonzero averages, and hence $\dot{S}^{\rm diss}(t) \geq 0$
at all times, in agreement with the second law of thermodynamics. This is explicitly shown in~\cite{lud16b} within an adiabatic expansion, valid for low driving frequencies. Here, we analyze an example beyond the adiabatic regime and show that $\dot{S}^{\rm diss}(t) \geq 0$ even though for some times, the net heat flow at the reservoirs might be negative, $\sum_{\alpha} \dot{Q}_{\alpha}(t) <0$.

\section{Calculation of the Currents within Green's Function Approach for Non Interacting Systems}

We now consider that the central part of the system is composed by a discrete chain of $N$ sites (e.g., an array of quantum dots), to which many time-periodic potentials $V_i(t)$ are applied. Then, the Hamiltonian of the central piece reads:
\beq
H_S(t)=H_S^0+\sum_m V_m(t)d_m^\dagger d_m,
\edq
where
\beq
H_S^0=\sum_{m=1}^N \veps_0^m d_m^\dagger d_m+\sum_{m=1}^{N-1}\left(t_{m,m+1} d_m^\dagger d_{m+1}+h.c.\right)
\edq
is the stationary part of the Hamiltonian, with $\veps_0^m$ the energy of the site $m$ and $t_{m,m+1}$ the hopping between two consecutive sites. 

\subsection{Time Resolved Charge and Energy Currents Entering the Reservoirs}

The charge and energy fluxes are defined from the expectation value of the time derivative of the number of particles and the energy of the reservoir, as in Equations~(\ref{encur}) and~(\ref{charcur}).
These currents are positive when entering the reservoir $\alpha$ and since $[H_{res},N_\alpha]=[H_{S},N_\alpha]=[H_{res},H_{\alpha}]=[H_{S},H_{\alpha}]=0$, they can be written as:

\begin{equation}\label{charge}
I_\alpha(t)=\dfrac{ie}{\hbar}\langle[H_{c\alpha},N_\alpha]\rangle=-\dfrac{2e}{\hbar}{\rm Re}\left\lbrace \sum_{k_\alpha}{w_{k_\alpha } \text{ }G^{<}_{l_\alpha,k_\alpha}(t,t)} \right\rbrace \text{  ,}
\end{equation}
and
\begin{equation}\label{energy}
J_\alpha^E(t)=\dfrac{i}{\hbar}\langle[H_{c\alpha},H_{\alpha}]\rangle=-\dfrac{2}{\hbar}{\rm Re}\left\lbrace \sum_{k_\alpha}{\veps_{k_\alpha}w_{k_\alpha} \text{ }G^{<}_{l_\alpha,k_\alpha}(t,t)} \right\rbrace \text{  ,}
\end{equation}
where $G^{<}_{l_\alpha ,k_\alpha}(t,t)\equiv i\langle{c^{\dagger}}_{k_\alpha}(t)d_{l_\alpha}(t)\rangle$ is a Green's function that involves operators of the reservoir ${c^{\dagger}}_{k_\alpha}$, as well as the central part, $d_{l_\alpha}$. From the Dyson equation and Langreth rules~\cite{jauho}, Green's function above can be expressed as follows
\begin{equation}
\label{menor}
G^{<}_{l_\alpha,k_\alpha}(t,t)=\int_{-\infty}^{\infty}dt_1w_{k_\alpha }^{\ast}[G^{R}_{l_\alpha, l_\alpha}(t,t_1)g_{k_\alpha}^{<}(t_1-t)+G_{l_\alpha,l_\alpha}^{<}(t,t_1)g_{k_\alpha}^{A}(t_1-t)]\text{  ,}
\end{equation}
with $G_{m,n}^{R}(t,t')=-i\theta(t-t')\langle\{d_m(t)d_n^{\dagger}(t')\}\rangle$ and $G_{m,n}^{<}(t,t')=-i\langle\{d_n^{\dagger}(t')d_m(t)\}\rangle$ being the retarded and the lesser Green's functions of the central piece, and 
\begin{equation}\label{gr}
g_{k_\alpha}^{<}(t-t')=i\int_{-\infty}^{\infty}\dfrac{d\veps}{2\pi}f_\alpha(\veps)\rho_{k_\alpha}(\veps)e^{-i\veps(t-t')/\hbar}\text{  ,}
\end{equation}
\begin{equation}\label{ga}
g_{k_\alpha}^{A}(t-t')=\int_{-\infty}^{\infty}\dfrac{d\veps}{2\pi}\int_{-\infty}^{\infty}\dfrac{d\veps '}{2\pi}\dfrac{\rho_{k_\alpha}(\veps ')}{\veps -\veps ' - i0^{+}}e^{-i\veps(t-t')/\hbar}\text{  ,}
\end{equation}
the Green's functions of the uncoupled reservoirs, with $\rho_{k_\alpha}(\veps)=2\pi \delta (\veps -\varepsilon_{k_\alpha})$, and $f_\alpha(\veps)=[e^{(\veps-\mu_\alpha)/(k_B T)}+1]^{-1}$ the Fermi-Dirac distribution. Then, substituting Equations~(\ref{menor})--(\ref{ga}) in Equation~(\ref{charge}), the charge current entering the reservoir $\alpha$ can be written as
\beq\label{c2}
I_\alpha(t) = -\frac{2e}{h}{\rm Re}\left\{\int dt_1 \int{d\veps} e^{-i\veps(t_1-t)/\hbar}\left[ iG_{l_\alpha,l_\alpha}^{R}(t,t_1)f_\alpha(\veps)\Gamma_\alpha(\veps) +G_{l_\alpha,l_\alpha}^{<}(t,t_1)\int {\dfrac{d\veps'}{2\pi}\dfrac{\Gamma_\alpha(\veps')}{(\veps-\veps' -i0^+)}}\right]\right\},
\edq
where we have defined the hybridization as $\Gamma_\alpha=\sum_{k_\alpha}\vert w_{k_\alpha}\vert^2\rho_{k_\alpha}$. We now consider the Dyson equation for the lesser Green's function \cite{arr-mos}
\beq\label{glesser}
G_{n,m}^<(t,t')=\sum_\alpha\int dt_1 dt_2G_{n,l_\alpha}^R(t,t_1)\Sigma_\alpha^<(t_1-t_2)[G_{m,l_\alpha}^R(t',t_2)]^*,
\edq
with $\Sigma_\alpha^<(\veps)=if_\alpha(\veps)\Gamma_\alpha(\veps)$, and we introduce the Floquet-Fourier representation \cite{arrg1}, which is convenient for periodic fields:
\beq\label{floquet11}
\hat{G}^R(t,t')=\int\frac{d\veps}{2\pi}\hat{G}^R(t,\veps)e^{-i\veps(t-t')/\hbar},\edq
with
\beq\label{floquet2}
\hat{G}^R(t,\veps) =\sum_{n=-\infty}^{\infty}\hat{\calg}(n,\veps)e^{-in\omega t},
\edq
where $\omega=2\pi/\tau$ is the oscillation frequency of the AC parameters ${\bf{V}}(t)$, and $\hat{\calg}(n,\veps)$ are the Floquet components. Green's function $\hat{G}^R(t,\veps)$, can be obtained by solving the Dyson equation \cite{arrg1,arrg2}:
\beq\label{dyson}
\hat{G}^R(t,\veps)=\hat{G}^0(\veps)+\sum_{n=-\infty}^{\infty}\hat{G}^R(t,\veps+n\hbar\omega)\hat{V}(n)\hat{G}^0(\veps)e^{-in\omega t},
\edq
with $\hat{G}^0(\veps)=[\veps \hat{I}-\hat{H}_S^0-i\hat{\Gamma}/2]^{-1}$ being the stationary retarded Green's function and $\hat{\Gamma}_{m,n}=\sum_{\alpha=L,R}\delta_{m,n}\delta_{l_\alpha,m}\Gamma_\alpha$. The elements of the matrix $\hat{V}_{l,l'}(n)=\delta _{l,l'}\sum_i \delta_{l,i}V_i(n)$ are the Fourier components of $\hat{V}_{l,l'}(t)=\sum_n \hat{V}_{l,l'}(n)e^{in\omega t}$.

Replacing Equations~(\ref{floquet11}) and~(\ref{floquet2}) into Equation~(\ref{c2}), we get  
\ba
I_\alpha(t)& = & -\frac{2e}{h}{\rm Re}\left\{\sum\limits_{l}e^{-il\omega t} \int{d\veps}\left[ i{\mathcal{G}}_{l_\alpha,l_\alpha}(l,\veps)f_\alpha(\veps)\Gamma_\alpha(\veps)\right.\right.\\
& & \left.\left.+\sum\limits_{\beta=L,R}\sum\limits_{n}i{\mathcal{G}}_{l_\alpha,l_\beta}(l+n,\veps){{\mathcal{G}}}_{l_\alpha,l_\beta}^{\ast}(n,\veps)f_\beta(\veps)\Gamma_\beta(\veps)\int {\dfrac{d\veps'}{2\pi}\dfrac{\Gamma_\alpha(\veps ')}{(\veps-(\veps ' -n\hbar\omega) -i0^+)}}\right]\right\}.\nonumber
\ea

It is convenient to express the charge current in terms of differences between the Fermi functions of the reservoirs. To this end, we first explicitly write the real part as the sum of a complex number and its conjugate
\ba
I_\alpha(t) & = & -\frac{e}{h}\int {d\veps}\sum\limits_{l}ie^{-il\omega t}\left[f_\alpha(\veps)\Gamma_\alpha(\veps)\left(\mathcal{G}_{l_\alpha,l_\alpha}(l,\veps)-\mathcal{G}_{l_\alpha,l_\alpha}^{\ast}(-l,\veps) \right)\right.\\\nonumber
 & &  +\sum\limits_{\beta =L,R}\sum\limits_{n} f_\beta(\veps)\Gamma_\beta(\veps)\mathcal{G}_{l_\alpha,l_\beta}(l+n,\veps)\mathcal{G}_{l_\alpha,l_\beta}^{\ast}(l,\veps)\\\nonumber
 & & \left.\int\dfrac{d\veps '}{2\pi}\left(\dfrac{\Gamma_\alpha(\veps')}{\veps -(\veps '-n\hbar\omega)-i0^+}-\dfrac{\Gamma_\alpha(\veps'+l\hbar\omega)}{\veps -(\veps '-n\hbar\omega)+i0^+} \right)\right].
\ea
Within the wide band limit, the hybridization can be considered a constant function $\Gamma_\alpha(\veps)\sim \Gamma_\alpha$. Taking into account also the following relation
\begin{equation}
\label{bueno}
\dfrac{1}{\veps -\veps ' -i0^+}=\mathcal{P}\lbrace\dfrac{1}{\veps -\veps '}\rbrace +i\pi\delta(\veps -\veps '),
\end{equation}
we find
\ba
I_\alpha(t)& = & -\frac{e}{h}\int {d\veps}\sum\limits_{l}ie^{-il\omega t}\Gamma_\alpha\left[ f_\alpha(\veps)\left(\mathcal{G}_{l_\alpha,l_\alpha}(l,\veps)-\mathcal{G}^{\ast}_{l_\alpha,l_\alpha}(-l,\veps) \right)\right.\\\nonumber
& & \left.+i\sum\limits_{\beta =L,R}\sum\limits_{n}f_\beta(\veps -n\hbar\omega)\Gamma_\beta\mathcal{G}_{l_\alpha,l_\beta}(l+n,\veps -n\hbar\omega)\mathcal{G}_{l_\alpha,l_\beta}^{\ast}(n,\omega -n\omega)\right].
\ea
Finally, using the identity for the spectral components \cite{arr-mos}
\beq
i\calg_{l,l'}(n,\veps)-i\calg_{l,l'}^\dagger(-n,\veps +n\omega)=\sum_{\beta=L,R}\sum_{n'}\calg_{l,l_\beta}(n+n',\veps -n'\omega)\Gamma_\beta\calg^{*}_{l',l_\beta}(n',\veps -n'\omega),
\edq
the charge current reads
\ba\label{jefinal}
I_\alpha(t) & = & -\frac{e}{h} \int {d\veps}\sum\limits_{l}e^{-il\omega t}\Gamma_\alpha\left\{i\mathcal{G}^{\ast}_{l_\alpha,l_\alpha}(-l,\veps )\left[f_\alpha(\veps-l\hbar\omega)-f_\alpha(\veps ) \right]\right.\\\nonumber
& & -\sum\limits_{\beta =L,R}\sum\limits_{n}\Gamma_\beta\mathcal{G}_{l_\alpha,l_\beta}(l+n,\veps -n\hbar\omega)\mathcal{G}_{l_\alpha,l_\beta}^{\ast}(n,\veps-n\hbar\omega)\left[f_\beta(\veps-n\hbar\omega)-f_\alpha(\veps ) \right]\rbrace .
\ea

Following the same procedure, we can express the energy current entering reservoir $\alpha$ in Equation (\ref{energy}) as
\beq\label{ce2}
J_\alpha^E(t) = -\frac{2e}{h}{\rm Re}\left\{\int dt_1 \int{d\veps} e^{-i\veps(t_1-t)/\hbar}\left[ iG^{R}_{l_\alpha,l_\alpha}(t,t_1)f_\alpha(\veps)\Gamma_\alpha(\veps)\veps +G_{l_\alpha,l_\alpha}^{<}(t,t_1)\int {\dfrac{d\veps'}{2\pi}\dfrac{\Gamma_\alpha(\veps')\veps'}{(\veps-\veps' -i0^+)}}\right]\right\},
\edq
where we have used inside the integral that it is possible to make the replacement $\veps_{k_\alpha}\delta(\veps-\veps_{k_\alpha})\rightarrow\veps\,\delta(\veps-\veps_{k_\alpha})$. Now, using the Floquet representation for Green's functions and repeating the same steps as for the charge current, we find
\ba\label{jefinale}
J^E_\alpha(t) & = & -\frac{e}{h} \int {d\veps}\sum\limits_{l}e^{-il\omega t}\Gamma_\alpha\left\{i\mathcal{G}_{l_\alpha,l_\alpha}^{\ast}(-l,\veps )\left[(\veps-l\hbar\omega)f_\alpha(\veps-l\hbar\omega)-\veps f_\alpha(\veps ) \right]\right.\nonumber\\
& & -\sum\limits_{n,\beta =L,R}\sum\limits_{n}\Gamma_\beta\mathcal{G}_{l_\alpha,l_\beta}(l+n,\veps -n\hbar\omega)\mathcal{G}_{l_\alpha,l_\beta}^{\ast}(n,\veps-n\hbar\omega)\\
&&\left.\left[(\veps-\frac{l\hbar\omega}{2})f_\beta(\veps-n\hbar\omega)-\veps f_\alpha(\veps ) \right]\right\} \nonumber.
\ea

\subsection{Energy Stored in the Contact Regions}\label{apa}
The rate of change for the energy in the contact regions $J_{c\alpha}^E(t)=\langle \dot{H}_{c\alpha}\rangle $ can be also expressed in terms of Green's function $G^<_{l_\alpha,k_\alpha}(t,t)$ 
\beq
J_{c\alpha}^E(t)=\sum_{k_\alpha} \left[w_{k_\alpha}\frac{d\left< c_{k_\alpha}^{\dagger}(t)d_{l_\alpha}(t) \right>}{dt} + h.c\right]=-2\sum_{k_\alpha}{\mbox{Im}\left\{w_{k_\alpha}\frac{d{G}_{l_\alpha,k_\alpha}^{<}(t,t)}{dt}\right\}}.
\edq

Following a similar procedure for the charge and energy currents, we find that the energy stored in the contact region reads
\beq 
J_{c\alpha}^E(t)=\int\frac{d\veps}{2\pi}\omega {\Gamma}_\alpha
\sum_l l\left[2\mbox{Im}\{e^{-i l\omega t}{\cal G}_{l_\alpha,l_\alpha}(l,\veps)\}f_\alpha(\veps)+\sum_{n}\sum_{\beta=L,R}f_\beta(\veps){\Gamma}_\beta\mbox{Re}\{e^{-i l\omega t}{\cal{G}}_{l_\alpha,l_\beta}(l+n,\veps){{\cal G}}_{l_\alpha,l_\beta}^{*}(n,\veps)\}\right].
\edq
It is easy to show that the second term of the above equation vanishes, since:
\ba
\sum_{l>0}\sum_n\left(l\,\mbox{Re}\{e^{-i l\omega t}{\cal{G}}_{l_\alpha,l_\beta}(l+n,\veps){{\cal{G}}}_{l_\alpha,l_\beta}^{*}(n,\veps)\}-l\,\mbox{Re}\{e^{i l\omega t}{\cal{G}}_{l_\alpha,l_\beta}(-l+n,\veps){{\cal{G}}}_{l_\alpha,l_\beta}^{*}(n,\veps)\}\right)\nonumber
\ea
\ba
=\sum_{l>0}\sum_n\,l\,\left(\mbox{Re}\{e^{-i l\omega t}{\cal{G}}_{l_\alpha,l_\beta}(l+n,\veps){{\cal{G}}}^{*}_{l_\alpha,l_\beta}(n,\veps)\}-\mbox{Re}\{e^{i l\omega t}{{\cal{G}}}^{*}_{l_\alpha,l_\beta}(l+n,\veps){{\cal{G}}}_{l_\alpha,l_\beta}(n,\veps)\}\right)=0.\nonumber
\ea
Hence,  
\ba
\label{diff2}
J_{c\alpha}^E(t) 
& = & \int\frac{d\veps}{h} f_\alpha(\veps){\Gamma}_\alpha\sum_l l\hbar\omega\,2\mbox{Im}\{e^{-i l\omega t}{\cal{G}}_{l_\alpha,l_\alpha}(l,\veps)\}.
\ea
Combining this expression with Equation (\ref{floquet2}), we find the expression:
\ba
J_{c\alpha}^E(t) 
& = & \int\frac{d\veps}{2\pi} f_\alpha(\veps){\Gamma}_\alpha 2\mbox{Re}\{ {\partial_t {G}_{l_\alpha,l_\alpha}^R}(t,\veps)\},
\ea
which clearly shows that this energy reactance term is of purely an AC nature since it vanishes after time averaging.

\subsection{Power Developed by the AC Sources}

The power performed by the AC potential can be calculated as $P_{AC}(t)=-\left\langle\frac{\partial H_S}{\partial t}\right\rangle$, which is
\beq
P_{AC}(t)=-\sum_i\frac{dV_i(t)}{dt}\langle d_i^\dagger(t)d_i(t)\rangle=\sum_i\frac{dV_i(t)}{dt} \mbox{Im}\{G_{i,i}^<(t,t)\}.
\edq
Now, replacing Equation (\ref{glesser}) in the above equation , and using the Floquet representation, we get
\beq\label{powerfloq}
P_{AC}(t)=\sum_i\sum_{\alpha=L,R}\sum_{l,m,n}\int \frac{d\veps}{h}n\hbar\omega f_\alpha(\veps)\Gamma_\alpha \mbox{Im}\{V_i(n)\calg_{i,l_\alpha}(m+l,\veps)\calg_{i,l_\alpha}^*(l,\veps)e^{-i\omega t(m-n)}\}.
\edq

\section{Relation to the Scattering Matrix Formalism}\label{secscatt}

In this section, we show that the definition of the heat current in  {Equation (\ref{defheatres})}, which includes the contribution $J_{c\alpha}^E$ due to the contact $\alpha$,  is fully in agreement with the scattering matrix formalism. From Equations (\ref{jefinal}), (\ref{jefinale}) and (\ref{diff2}), the heat reads
\ba\label{heatfloq}
\dot{Q}_\alpha (t)&& =\sum_l\int\frac{d\veps}{h}e^{-il\omega t} \Gamma_\alpha \left\{ i\calg_{l_\alpha,l_\alpha}^*(-l,\veps)(\veps-\frac{l\hbar\omega}{2}-\mu_\alpha)(f_\alpha(\veps)-f_\alpha(\veps -l\hbar\omega))\right.\\
&&-\left.\sum_n\sum_{\beta=L,R} (\veps+\frac{l\hbar\omega}{2}-\mu_\alpha)(f_\alpha(\veps)-f_\beta(\veps -n\hbar\omega))\Gamma_\beta \calg_{l_\alpha,l_\beta}(l+n,\veps-n\hbar\omega)\calg_{l_\alpha,l_\beta}^*(n,\veps-n\hbar\omega)\right\}\nonumber
\ea

Within  the scattering matrix approach, the heat flux entering the reservoir $\alpha$ reads \cite{lim13,batt-mos}
\beq
\dot{Q}^S_\alpha (t)=\sum_{l,n} e^{-il\omega t}\int\frac{d\veps}{h}(\veps+\frac{l\hbar\omega}{2}-\mu_\alpha)\sum_{\beta=L,R}(f_\beta(\veps_{-n})-f_\alpha(\veps))S^*_{\alpha\beta}(\veps,\veps_{-n})S_{\alpha,\beta}(\veps_l,\veps_{-n}),
\edq 
where $\veps_n=\veps+n\hbar\omega$ and $S(\veps_n,\veps_l)$ is the Floquet scattering matrix which is related to the Green function via the generalized Fisher-Lee relation \cite{arr-mos}
\beq\label{paso}
S_{\alpha\beta}(\veps_m,\veps_n)=\delta_{\alpha,\beta}\delta_{m,n}-i\sqrt{\Gamma_\alpha(\veps_m)\Gamma_\beta(\veps_n)}\mathcal{G}_{l_\alpha,l_\beta}(m-n,\veps_n).
\edq
Using this relation, we find that
\ba
\dot{Q}^S_\alpha (t)&=&\sum_{l,n}\int\frac{d\veps}{h}e^{-il\omega t}(\veps+\frac{l\hbar\omega}{2}-\mu_\alpha) \sum_{\beta=L,R} (f_\beta(\veps -n\hbar\omega)-f_\alpha(\veps))\\
&&\times\calg_{l_\alpha,l_\beta}^*(n,\veps-n\hbar\omega)\bigg\{i\delta_{\alpha\beta}\delta_{l,-n}\sqrt{\Gamma_\alpha\Gamma_\beta}+\Gamma_\alpha\Gamma_\beta\calg_{l_\alpha,l_\beta}(l+n,\veps-n\hbar\omega)\bigg\},\nonumber
\ea
Then, after some algebra and by comparing with Equation (\ref{heatfloq}), we have $\dot{Q}^S_\alpha (t)=\dot{Q}_\alpha (t)$.
\section{Low Frequency Expansion}\label{lfrec}
For low frequencies of the driving potentials, a solution of the Dyson equation (\ref{dyson}) up to ${\cal{O}}(\omega)$ can be obtained by expanding in powers of the AC frequency as
\ba 
\hat{G}^R(t,\veps) & \sim &\hat{G}^{0}(\veps) +
\hat{G}^R(t,\veps) \hat{V}(t) \hat{G}^{0}(\veps) + i \hbar\partial_{\veps} \hat{G}^R(t,\veps) \frac{d \hat{ V}(t)}{dt}
\hat{G}^{0}(\veps).
\ea
Now, we can define the frozen Green's function
\beq\label{froz}
\hat{G}^R_f (t,\veps) = \left[ \hat{G}^{0}(\veps)^{-1} - \hat{V}(t)
\right]^{-1},
\edq
in terms of which the exact solution of Equation~(\ref{dyson}) up to first order in $\omega$ reads
\beq\label{gfrozen}
\hat{G}^{R}(t,\veps) = \hat{G}^{R}_f (t,\veps) + \frac{i\hbar}{2} \left(\frac{\partial^2}{\partial t \partial \veps} \hat{G}^{R}_f (t,\veps)+\hat{A}(t,\veps)\right),
\edq
with
\beq
\hat{A}(t,\veps)=\partial_\veps\hat{G}^R_f(t,\veps)\hat{V}(t)\hat{G}^R_f(t,\veps)-\hat{G}^R_f(t,\veps)\hat{V}(t)\partial_\veps\hat{G}^R_f(t,\veps).
\edq
Note that this expression is valid 
to lowest order in $\omega$, but can be used for arbitrary
values of the amplitude $V_{AC}$~\cite{mos-bu2009,arr-mos}. 
This is specially appealing
if one is interested in addressing nonlinear effects~\cite{mos08,kas12,alo16}.
The validity of the adiabatic approximation considered here was confirmed in~\cite{noc16} by a comparison with an exact calculation 
in the limit of low level occupations. 

We can also expand the Floquet components $\hat{\calg}(n,\veps)$ up to first order in $\omega$ as
\beq
\hat{{\cal G}}(n,\veps)\sim\hat{\cal G}^{(0)}(n,\veps)+\hbar\omega\hat{\cal G}^{(1)}(n,\veps),
\edq
Then, using the representation of Equation~(\ref{floquet2}) in Equation~(\ref{gfrozen}), it is possible to identify the first and second terms of the expansion:
\ba
\hat{\cal G}^{(0)}(n,\veps)& = & \int_{0}^{\tau}\frac{dt}{\tau}\hat{G}^{R}_f(t,\veps)e^{in\omega t}\nonumber\\
\omega \hat{\cal G}^{(1)}(n,\veps)& = & \int_{0}^{\tau}\frac{dt}{\tau}\frac{i}{2}\left(\frac{\partial^2}{\partial t \partial \veps} \hat{G}^{R}_f (t,\veps)+\hat{A}(t,\veps)\right)e^{in\omega t}.
\ea

\section{Application}
We now illustrate our main ideas with a simple but generic example: a single driven level coupled
to two fermionic baths kept at the same temperature, but with different chemical potentials, $\mu_L=\mu$ and $\mu_R=\mu-\delta \mu$. The Hamiltonian of the sample is then
\beq
\calh_S=\veps_d(t)d^\dagger d\,,
\edq
where the level is periodically driven with a monochromatic potential of frequency $\omega$ and
amplitude $V_{\rm AC}$, $\veps_d(t)=\veps_0+V(t)$ with $V(t)=V_{\rm AC}\cos\omega t$.
The response of the system will be periodic.
Another possibility is to explore heat dynamics in the transient regime, usually in quantum
levels whose potential is suddenly changed by an externally-applied voltage pulse.
Thermoelectric efficiency can thus be enhanced by a suitable time-dependent potential applied to
a quantum dot level far from equilibrium~\cite{cre11,zho15}.
Based on a nonequilibrium Green's function approach,~\cite{yu14}
investigates the transient heat current using a complex absorbing potential,
greatly reducing the complex calculation of the heat current through a molecule connected
to metallic leads and under a gate voltage pulse. Indeed,
the transient heat current of a quantum dot in response to a voltage shift
can reveal signatures of electronic interaction despite the inherent Coulomb repulsion~\cite{sch16}.
The full counting statistics of the quantum energy flow can also be determined for the transient regime,
showing universal scaling laws in the single quantum dot case~\cite{yu16}.
When the system consists of two levels coupled to a thermal bath, the transient dynamics
is sensitive to the employed theoretical formalism (master equation approach or functional
influence method)~\cite{car16}.

Here, we restrict ourselves to mean currents. However, their fluctuations
can provide us with useful information. 
For instance, fluctuations of heat beyond the level dictated by the  Callen-Welton fluctuation-dissipation theorem signal the presence of backscattering~\cite{mos14}. 
Furthermore, the energy noise reveals additional information
not contained in the charge current fluctuations~\cite{bat14}.
The novel features are related to electron-hole correlations arising in phase-coherent
small conductors driven by periodic voltage biases.
The mixed noise between correlations of charge and energy currents
shows both interference and transport contributions, and is unique as compared
with either charge or energy noises~\cite{san13,bat14b,cre15,cre16,eym16}.


\subsection{Adiabatic Regime and Linear Response in the Bias Voltage}

For small bias voltage, it is possible to combine the procedure described in Section~\ref{lfrec} with an expansion in $\delta\mu$. To calculate up to second order in $\omega$ and $\delta\mu$ the heat current of Equation~(\ref{heatfloq}), the AC power in Equation~(\ref{powerfloq}), and the power developed by the batteries in Equation~(\ref{powerbat}), we need to perform an expansion of the Fermi functions entering the integrals as:
\beq
f_\alpha(\veps+n\hbar\omega)\simeq f_\alpha(\veps)+\partial_\veps f_\alpha n\hbar\omega +\partial^2_\veps f_\alpha\frac{(n\hbar\omega)^2}{2},
\edq
with 
\ba
f_L(\veps)&\equiv & f(\veps)\\
f_R(\veps)&\simeq & f(\veps)+\partial_\veps f\delta\mu+\partial^2_\veps f\frac{\delta\mu ^2}{2}.
\ea

Accordingly, we can compute all of the fluxes entering Equation~(\ref{2law-1}). Then, up to second order in $\omega$ and $\delta\mu$, the total dissipative power $P_{\rm tot}^{\rm diss}$ reads:
\beq\label{Ppower}
P_{\rm tot}^{\rm diss}(t)=L^{\delta\mu^2}(t)\delta\mu^2+L^{\omega^2}(t)\omega^2,
\edq
where the transport coefficients are:
\ba\label{lambdas}
L^{\delta\mu^2}(t)&=&\int\frac{d\veps}{h}\partial_\veps f \Gamma_L\Gamma_R\vert G^R_f(t,\veps)\vert^2\\
L^{\omega^2}(t)&=&\frac{\hbar}{2\omega^2}\int\frac{d\veps}{2\pi}\partial_\veps f {\Gamma}^2\vert\partial_tG^R_f(t,\veps)\vert^2,\nonumber
\ea
where $\Gamma=\Gamma_L+\Gamma_R$ is the total hybridization.
Notice that, as is shown is~\cite{lud16} for  systems with time-reversal symmetry, the terms that are proportional to $\omega\delta\mu$ do not contribute to the total dissipated power . 

Following a similar procedure in Equation~(\ref{heatfloq}) and then subtracting Equation~(\ref{Ppower}), we find that the dissipated heat flux of the central piece is
\beq
\dot{Q}^{\rm diss}_S(t)=L_d^{\omega}(t)\omega+L_d^{\omega\delta\mu}(t)\omega\delta\mu+L_d^{\omega^2}(t)\omega^2\,,
\edq
with
\ba
& &L_d^{\omega}(t)=-\int \frac{d\veps}{2\pi}\partial_\veps f (\veps-\mu)\Gamma\vert G^R_f(t,\veps)\vert^2 V_{\rm ac}\mbox{sin}(\omega t)\,,\nonumber\\
& &L_d^{\omega\delta\mu}(t)=\int \frac{d\veps}{2\pi}\Gamma_R\partial_\veps f (\veps-\mu)\partial_\veps\vert G^R_f(t,\veps)\vert^2 V_{\rm AC}\mbox{sin}(\omega t)\,, \\
& &L_d^{\omega^2}(t)=\frac{\hbar}{2\omega^2}\int \frac{d\veps}{2\pi}\Gamma^2\partial_\veps f (\veps-\mu)\partial_t(\partial_\veps G^R_f(t,\veps)\partial_t {G^R_f}^*(t,\veps)).\nonumber
\ea


\begin{figure}[ht!]
\centering
\includegraphics[width=8cm]{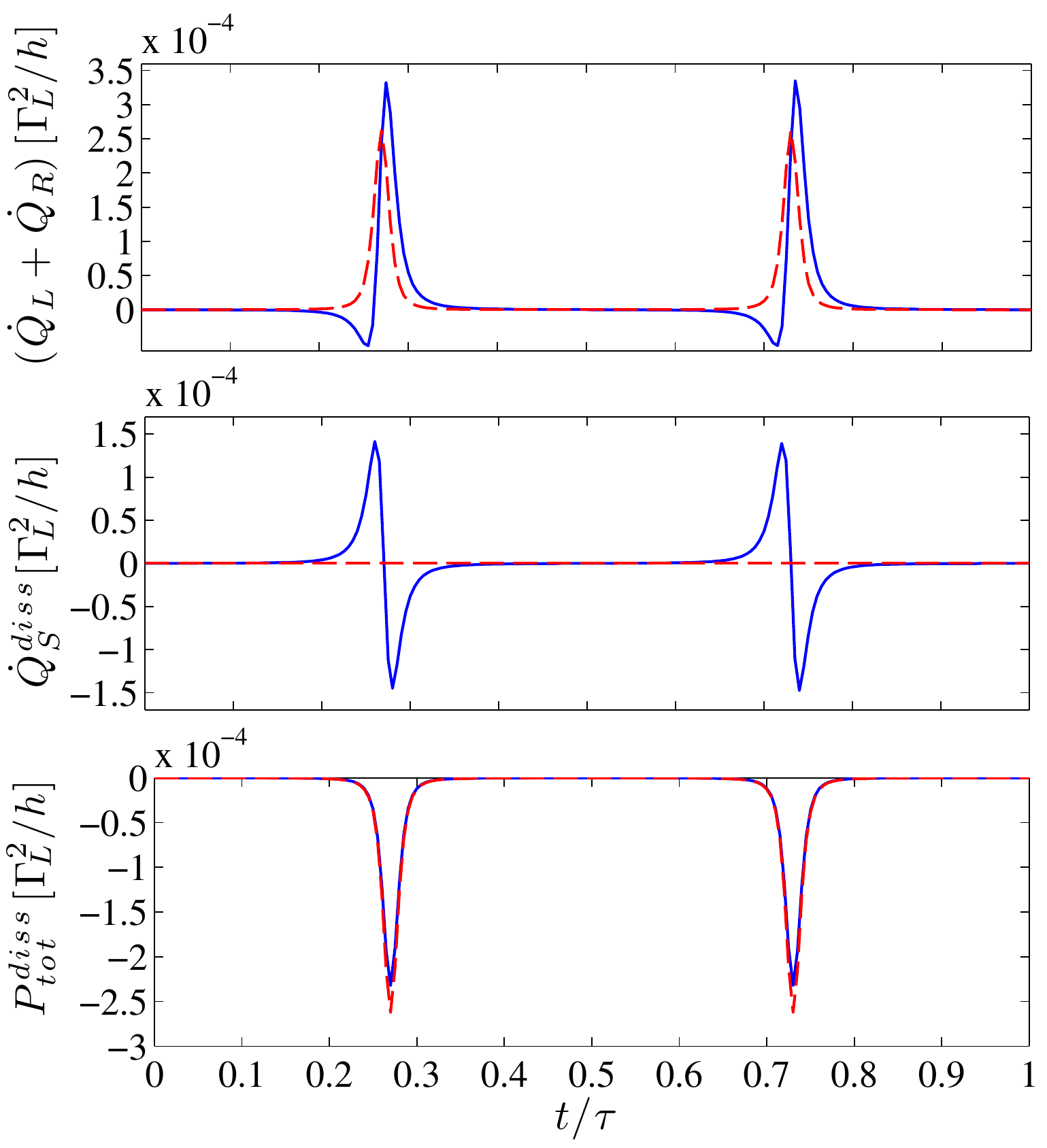}
\caption{Adiabatic regime of a single driven quantum level coupled to two fermionic baths. The different components of the total heat production $Q^{\rm diss}_{\rm tot}(t)=-P^{\rm diss}_{\rm tot}(t)$ as a function of time. Dashed lines correspond to reservoirs at $T=0$, while solid lines are for $k_BT=0.02$. Energies are expressed in units of $\Gamma_L$. The energy of the level evolves in time with $V_{AC}=-8$ and $\hbar\omega=1\times 10^{-3}$. Parameters: $\mu=1$, $\delta\mu=0.004$, $\veps_0=0$ and hybridization widths $\Gamma_L =1$ and $\Gamma_R = 0.5$.
\label{fig_ad}}
\end{figure}   

In Figure~\ref{fig_ad} (upper panel) we show the total heat generated in the reservoirs as a function of time
for zero temperature (red dashed line) and nonzero temperature (blue line).
In the latter case, an individual heat flux can become negative but the total dissipated heat must be positive
for all temperatures, as shown in Figure~\ref{fig_ad} (lower panel). Note that one should also take into account the contribution
from the sample, which is purely dynamical [Figure~\ref{fig_ad} (middle panel)]
and vanishes for zero temperature.

\subsection{Nonadiabatic Regime}

We now consider the entropy production for high driving frequencies,
where the linear response approach of Section 9.1 is not applicable. The evaluation of the currents implies the evaluation of the exact Dyson equation 
(\ref{dyson}) following the methods of  \cite{arrg1,arrg2}. A similar situation was analyzed in~\cite{man15}, where it was found that the rate of change for the nonadiabatic energy agrees
with the work done on a molecule subjected to a time-dependent applied field.
The work in~\cite{gloria} discusses heat transfer across a triple dot structure with the nonadiabatic driving
applied to an edge dot.

\begin{figure}[H]
\centering
\includegraphics[width=8cm]{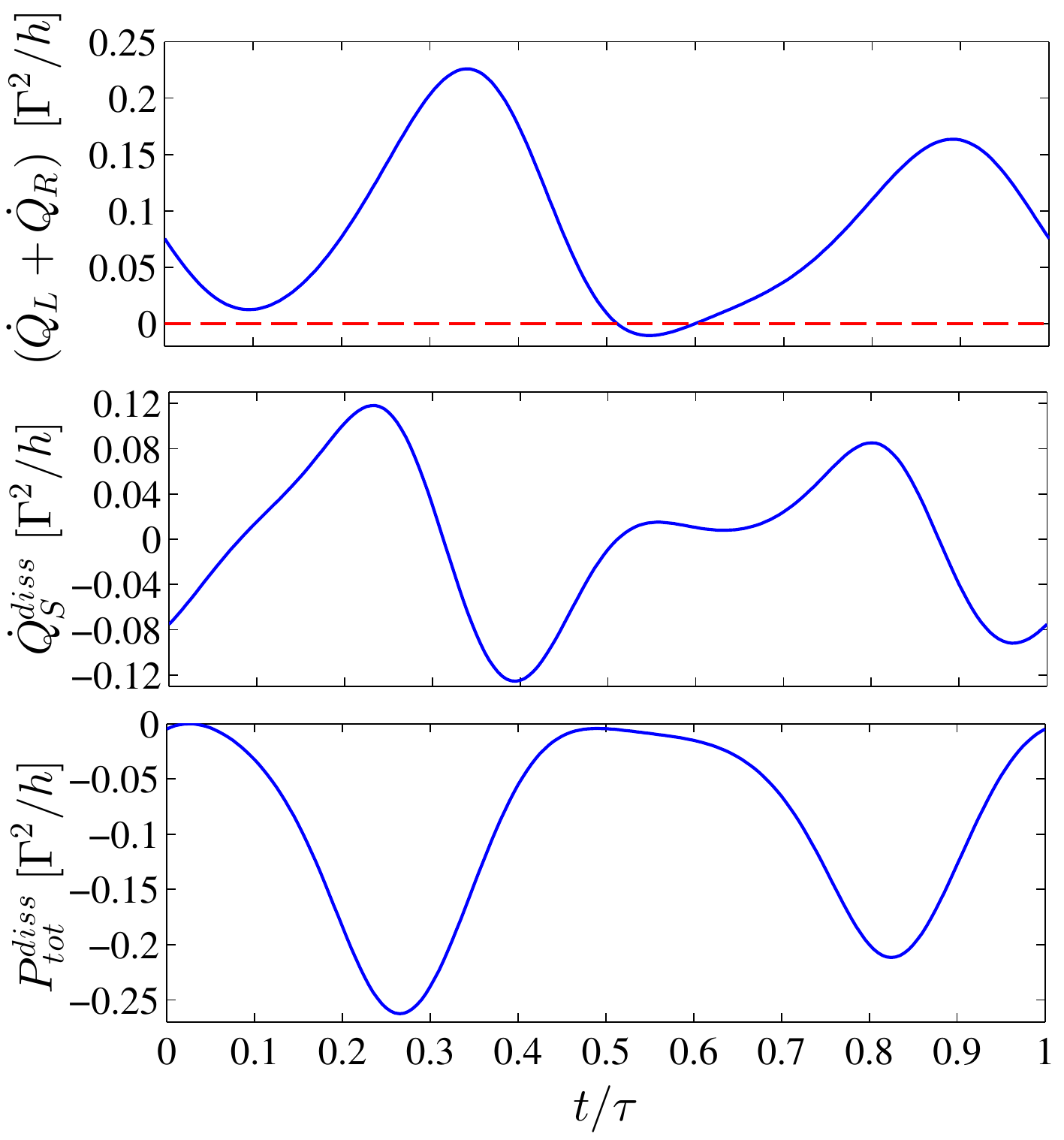}
\caption{Nonadiabatic regime of a single driven quantum level coupled to two fermionic baths at $T=0$. The different components of the total heat production $Q^{\rm diss}_{\rm tot}(t)=-P^{\rm diss}_{\rm tot}(t)$ as a function of time. The energy of the level evolves in time with $V_{AC}=0.7$ and $\hbar\omega=0.3$. Parameters: \protect\linebreak $\mu=0.2$, $\delta\mu=0$, $\veps_0=0$, and hybridization widths $\Gamma_L =\Gamma_R = 0.5$. Energies are expressed in units of $\Gamma=\Gamma_L+\Gamma_R$. The upper panel shows that the heat flux at the reservoirs may instantaneously attain negative values. The dissipative heat flux at the driven dot $Q^{\rm diss}_S(t)$ is highly fluctuating but the sum of the two contributions satisfies $Q^{\rm diss}_{\rm tot}(t)\geq 0$, consistent with the second law.
\label{fig_nad}}
\end{figure}

In our case, we focus on reservoirs at zero temperature for simplicity.
We see in the figure that, unlike the adiabatic case, the heat flux
at the reservoirs may attain negative values
for short (but finite) time intervals  [see Figure~\ref{fig_nad} (upper panel)].
Nevertheless, we emphasize that the total
entropy production is always positive at every instant
[see Figure~\ref{fig_nad} (lower panel)]. For this to occur the dissipated heat
at the sample must be nonvanishing and positive [see Figure~\ref{fig_nad} (middle panel)].


\section{Conclusions}
The subject of time-dependent heat in quantum electronic systems presents unique aspects as compared with the better understood macroscopic case. In this work, we have briefly reviewed the recent developments in the field.
On the basis of general thermodynamic arguments and conservation laws 
we have presented exact equations for the time-resolved energy conversion in a quantum system in contact with AC driving sources and reservoirs at which
DC bias voltages are applied. These equations lead to the unambiguous definition of the total heat and production in the full system containing 
the central driven device, the reservoirs and the contacts between the different parts. These results are independent of the particular method used to evaluate the physical quantities involved. In order to get the exact solution of an example we have used, for convenience, a Green function technique as in~\cite{lud14,lud16b}. Of course, the
physical picture is beyond the method employed to solve the
particular problem.

We have also analyzed the possibility of defining the heat flows not only resolved in time but also in space, in the sense of identifying flows in
the reservoirs and through the piece that is in contact with the AC sources.
How to properly perform this split in a physically-meaningful way is a non-trivial task. Intuitively, we expect that deep in the reservoirs, we should have an equilibrium system. For adiabatically-driven systems, we have found that this seems to be, indeed the case. Then, by suitably generalizing 
the definition of the time-dependent heat current. We argue that the proper generalization is given by Equation (\ref{defheatres}), which implies taking into account the rate of change of the energy stored at the reservoir plus half of the one stored at the contact. Such a definition leads to results compatible with scattering matrix expressions for continuum models, and to the derivation of a Joule law when the reservoirs are at zero temperature.

However, for highly non-equilibrium driving at the central piece and 
strongly-coupled reservoirs, it is not obvious at all if we should ever recover an equilibrium state at every instant, even deep inside the reservoirs. In fact, we have analyzed the example of a single-level system driven by a high-frequency gate voltage in contact with reservoirs at zero temperature and
we have found that the definition of the instantaneous heat current at the reservoirs  [Equation~(\ref{defheatres})] 
that leads to meaningful physical results in the adiabatic regime may be negative at some instants, meaning a temporary flow of heat exiting the reservoirs. This result is accompanied by a finite value of what we have identified as the heat flow in the central piece. The sum of both contributions leads to a net instantaneous positive value of the total entropy production. Hence, the result  is in agreement with the second law, albeit non-intuitive. This is because the natural expectation is to have
a dissipative energy flow only into the reservoirs and not in the central device. This is, in fact, the case when we consider the average over one period of the driving. However, instantaneously, and for non-equilibrium
situations, the exact equation for the entropy production is given by Equation~(\ref{2law-1}),
which contains contributions associated to the reservoirs, contacts and the central device. The second law implies that the sum of all these contributions should 
be positive but no fundamental law imposes independent conditions on the individual terms.
Finally, we have commented on the difficulties of accessing experimentally a reliable test of the time- and spatially-resolved heat transport discussed in the present work.

\vspace{6pt} 


\acknowledgments{This work was supported by MINECO under Grant No.~FIS2014-52564, UBACyT, CONICET and MINCyT, Argentina.}

\authorcontributions{All the authors conceived the theoretical model and performed the calculations. They analyzed the results and wrote the paper. All the authors read and approved the final manuscript.}

\conflictofinterests{The authors declare no conflict of interest. } 

\bibliographystyle{mdpi}

\renewcommand\bibname{References}



\end{document}